\documentclass[twoside,11pt]{article}
\usepackage{jair, theapa, rawfonts}
\usepackage{graphicx}
\usepackage{letltxmacro}
\usepackage{url}
\usepackage{booktabs}
\usepackage{xspace}

\ShortHeadings{Models of Level-0 Behavior}
{Wright \& Leyton-Brown}

\usepackage{dsfont}

\providecommand{\R}{\mathds{R}}\renewcommand{\R}{\mathds{R}}

\usepackage{amsmath, amsthm, amssymb}
\theoremstyle{definition}
\newtheorem{definition}{Definition}

\DeclareMathOperator*{\argmin}{arg\,min}
\DeclareMathOperator*{\argmax}{arg\,max}

\newcommand{\noun}[1]{\textsc{#1}}

\LetLtxMacro{\citep}{\cite}
\LetLtxMacro{\citet}{\citeA}
\newcommand{\citeeg}[1]{\cite<e.g.,>{#1}}
\providecommand{\chapter}[1]{}
\providecommand{\paper}{paper\xspace} 

\newcommand{\journalonly}[1]{#1}
\newcommand{\thesisonly}[1]{}

\usepackage[disable]{todonotes} 
\newcommand{\jrw}[1]{\todo[inline,color=red!50]{JRW: #1}}
\newcommand{\klb}[1]{\todo[inline,color=green!50]{KLB: #1}}

\begin{document}

\title{Models of Level-0 Behavior for\\Predicting Human Behavior in Games\thanks{An early version of this work was published as \citep{wright14level}.}}

\author{\name James R.\ Wright \email jrwright@cs.ubc.ca \\
       \name Kevin Leyton-Brown \email kevinlb@cs.ubc.ca \\
       \addr Computer Science Department, University of British Columbia,\\
       Vancouver, BC, Canada\quad V6T 1Z4
       }


\maketitle

\begin{abstract}
  Behavioral game theory seeks to describe the way actual people (as compared to idealized, ``rational'' agents) act in strategic situations. Our own recent work has identified iterative models (such as quantal cognitive hierarchy) as the state of the art for predicting human play in unrepeated, simultaneous-move games \citep{wright12,wright16}.  Iterative models predict that agents reason iteratively about their opponents, building up from a specification of nonstrategic behavior called level-$0$.
  The modeler is in principle free to choose any description of level-$0$ behavior that makes sense for the given setting. However, in practice almost all existing work specifies this behavior as a uniform distribution over actions.  In most games it is not plausible that even nonstrategic agents would choose an action uniformly at random, nor that other agents would expect them to do so.  A more accurate model for level-$0$ behavior has the potential to dramatically improve predictions of human behavior, since a substantial fraction of agents may play level-$0$ strategies directly, and furthermore since iterative models ground all higher-level strategies in responses to the level-$0$ strategy.  Our work considers models of the way in which level-$0$ agents construct a probability distribution over actions, given an arbitrary game.
  Using a Bayesian optimization package called SMAC \citep{hutter10,hutter11,hutter12}, we systematically evaluated a large space of such models,
  each of which makes its prediction based only on general features that can be computed from any normal form game.
  In the end, we recommend a model that achieved excellent performance across the board: a linear weighting of features that requires the estimation of four weights.
  We evaluated the effects of combining this new level-$0$ model with several iterative models, and observed large improvements in the models' predictive accuracies.
\end{abstract}

\newcommand{\linear}[1]{{\small\texttt{linear#1}}}
\newcommand{\smac}{{\small\texttt{smac}}}

\chapter{Models of Level-0 Behavior}
\label{ch:level0}

\section{Introduction}
\thesisonly{
In quantal cognitive hierarchy models, such as the Spike-Poisson QCH model from Chapter~\ref{ch:variations}, agents do not reason arbitrarily deeply about their opponents' beliefs about beliefs about beliefs. Instead, they start from a simple nonstrategic strategy\footnote{In this work, we refer to agents that form explicit beliefs about the 
behavior of other agents as ``strategic,'' and agents that do not reason about other agents in this way as ``nonstrategic''.  Nonstrategic should not be taken as a synonym for unsophisticated or thoughtless.  Some of the level-$0$ behavior that we describe below is rather sophisticated.}
(the \emph{level-$0$} behavior), and then reason for some fixed number of iterations about responses to that strategy (e.g., a \emph{level-$2$} agent quantally best responds to the combined behaviors of level-$1$ and level-$0$ agents).
}

\journalonly{
It is well known that the standard game-theoretic assumption that agents will adopt \emph{Nash equilibrium} strategies---where each agent simultaneously responds optimally to all the others---is often a poor predictor of actual human behavior \citeeg{goeree01}. This is a particular problem for researchers building artificially intelligent systems to interact with humans in strategic settings, such as randomized security protocols \citeeg{pita08,jain10,yin12,Yang13a} or negotiation \citeeg{gal07,wolpert13}.
The field of \emph{behavioral game theory} aims to develop models that more accurately describe human strategic behavior, as evaluated using experimental data \citep{camerer03}.  Our own recent work has identified one particular model, \emph{quantal cognitive hierarchy}---an extension of the cognitive hierarchy model of \citet{camerer04cognitive}---as the state of the art behavioral model for predicting human play in unrepeated, simultaneous-move games \citep{wright12,wright16}.
%
The quantal cognitive hierarchy (QCH) model has two key components.  The first component is \emph{quantal response}; that is, agents respond stochastically to their incentives---playing high utility actions with high probability and low utility actions with low probability---rather than best responding.  This expresses the intuition that two actions that yield roughly equal utilities have a roughly equal chance of being chosen.  The second component is \emph{iterative reasoning}; that is, agents do not reason arbitrarily deeply about their opponents' beliefs about beliefs about beliefs, but instead start from a simple nonstrategic\footnote{We refer to agents that form explicit beliefs about the 
behavior of other agents as ``strategic'' and the alternative as ``nonstrategic''.  Nonstrategic should thus not be taken as a synonym for unsophisticated or thoughtless; some of the level-$0$ behavior that we describe in what follows is rather sophisticated.}
(\emph{level-$0$}) behavior, and then reason for some fixed number of iterations about responses to that starting point (e.g., a \emph{level-$2$} agent quantally best responds to the combined behaviors of level-$1$ and level-$0$ agents).
}

Thus, in order to make use of a quantal cognitive hierarchy model one must first commit to a specification of level-$0$ behavior. Indeed, this is true of iterative models in general, such as cognitive hierarchy \citep{camerer04cognitive} and level-$k$ \citep{stahl94,nagel95,costagomes01}. It is important to get this specification right, for two reasons. First, there is growing evidence that a substantial fraction of human players do act nonstrategically
\thesisonly{[\citealp{burchardi12}; Chapter~\ref{ch:parameter-analysis} of this dissertation].}
\journalonly{\citeeg{burchardi12}.}
Second, a level-$0$ model also drives predictions that are made about strategic agents: higher-level agents are assumed to act by responding strategically to lower-level agents' behavior. 

Almost all work in the literature that uses iterative models adopts the specification that level-$0$ agents play a uniform distribution over actions. (In Section~\ref{sec:l0-related} we discuss the few exceptions of which we are aware, each of which is based on an explicitly encoded intuition about a specific setting of interest.)
The uniform-distribution approach has the advantage that it does not require insight into a game's structure, and hence can be applied to any game. However, in many games it is not plausible that an agent would choose an action uniformly at random, nor that any other agent would expect them to do so. For example, consider a dominated action that always yields very low payoffs for all players.

In this \paper{} we consider the question of how to do better. Specifically, we investigate general rules that can be used to induce a level-$0$ specification from the normal-form description of an arbitrary game. 
\journalonly{
%
In the next section we formally define our setting, and describe the data, methods, and model that we used in our work.
In Section~\ref{sec:l0-models} we define the space of richer level-$0$ models that we searched.  We did this in two different ways: using a manual forward-selection procedure, and using a Bayesian optimization procedure.
We define these procedures and analyze the models that they selected in Section~\ref{sec:model-selection}.
Notably, we found that the richer level-$0$ models improved performance in the all of the iterative models that we considered, even though they were selected based only their impact on the performance of the Spike-Poisson QCH model.
We then briefly survey some related work in Section~\ref{sec:l0-related} before concluding in Section~\ref{sec:conclusions}.
}

\journalonly{

\section{Framework}
\label{sec:framework}

We focus on unrepeated, simultaneous-move \emph{normal-form games}.  Formally, a normal-form game $G$ is a tuple $(N,A,u)$, where $N$ is a finite set of \emph{agents}; $A=\prod_{i\in N}A_i$ is the set of possible \emph{action profiles}; $A_i$ is the finite set of \emph{actions} available to agent $i$;  $u=\{u_i\}_{i \in N}$ is a set of \emph{utility functions} $u_i:A \to \R$, each of which maps from an action profile to a utility for agent $i$. Let $\Delta(X)$ denote the set of probability distributions over a finite set $X$. Overloading notation, we represent the expected utility of a profile of mixed strategies $s \in S = \prod_{i \in N}\Delta(A_i)$ by $u_i(s)$.  By $a_{-i}$ we refer to the joint actions of all agents except  $i$.
%
Our objective is to find a behavioral model that maps from a game $G$ and a (human) agent $i$ to a probability distribution over $i$'s action set $A_i$ that predicts $i$'s behavior in $G$.  In this section we first describe a model that we have shown achieves state-of-the-art performance in predicting human behavior, Spike-Poisson QCH \citep{wright16}.
We then describe the dataset and methods that we used to learn parameters and evaluate the performance of extensions to Spike-Poisson QCH.

\subsection{Quantal Cognitive Hierarchy}
\label{sec:qch}
The first key component of the quantal cognitive hierarchy model is quantal best response, in which errors become less likely as they become more costly.
Like most of the behavioral game theory literature, we use the logit specification of this concept.
\begin{definition}[Quantal best response]
  Let $u_i(a_i,s_{-i})$ be agent $i$'s expected utility when playing action
  $a_i$ against strategy profile $s_{-i}$.  Then a \emph{quantal
  best response} $QBR_i(s_{-i};\lambda)$ by agent $i$ to $s_{-i}$ is a mixed
  strategy $s_i$ such that
  \begin{equation}
  s_i(a_i) =  \frac{\exp[\lambda\cdot u_i(a_i, s_{-i})]}{\sum_{a'_i}\exp[\lambda\cdot u_i(a'_i, s_{-i})]},\label{eq:qbr}
  \end{equation}
  where $\lambda$ (the \emph{precision}) represents agents' sensitivity to utility differences.\qed
\end{definition}
Unlike (classical) best response, which is a set-valued function,
quantal best response always returns a single mixed strategy.  When
$\lambda=0$, quantal response mixes uniformly over all of the agents' actions;
as $\lambda\to\infty$, quantal best response approaches best response.

The second key component of the quantal cognitive hierarchy model is iterative response, in which higher-level agents reason about and respond to lower-level agents.  Describing the distribution of different levels in the population of agents is a crucial decision.
We use the Spike-Poisson specification of this distribution \citep{wright16}

\begin{definition}[Spike-Poisson QCH model]
\label{def:qch} Let $\text{Poisson}(m;\tau)$ denote a Poisson distribution with mean $\tau$ evaluated at index $m$. Let
\[g(m) = \begin{cases}
           \epsilon + (1-\epsilon)\text{Poisson}(m;\tau) &\text{if } m=0, \\
           (1-\epsilon)\text{Poisson}(m;\tau)            &\text{otherwise.}
         \end{cases}\]
Let $QBR_i(s_{-i};\lambda)$ denote $i$'s quantal best response to the strategy profile $s_{-i}$, given precision parameter $\lambda$.  Let \[\pi_{i,0:m}=\sum_{\ell=0}^mg(\ell)\frac{\pi_{i,\ell}}{\sum_{\ell'=0}^mg(\ell')}\] be the truncated distribution over actions predicted for an agent conditional on that agent's having level $0 \le \ell \le m$.  Now we can define $\pi_{i,m} \in \Delta(A_i)$, the distribution over actions predicted for an agent $i$ with level $m$ by the Spike-Poisson QCH model:
  \begin{align*}
    \pi^\varnothing_{i,0}(a_i) &= |A_i|^{-1} \\
    \pi_{i,m}(a_i) &= QBR_i(\pi_{i,0:m-1};\lambda).
  \end{align*}
  Notice that $\pi^\varnothing_{i,0}$ is just the uniform distribution; that is, level-$0$ agents randomize uniformly among their actions.
  In Sections~\ref{sec:l0-models} and~\ref{sec:model-selection}, we will construct more plausible definitions of level-$0$ behavior by replacing $\pi^\varnothing_{i,0}$ with richer distributions.
  This baseline model thus has three parameters: the mean of the Poisson distribution $\tau$, the spike probability $\epsilon$, and the precision $\lambda$.
  The overall predicted distribution of actions is a weighted sum of the distributions for each level:
  \[\Pr(a_i \| G, \tau, \epsilon, \lambda) = \sum_{\ell=0}^\infty g(\ell)\pi_{i,\ell}(a_i).\hfill\qed\]
\end{definition}

In previous work, we found that the distributions of levels best supported by experimental data tended to have a large proportion of level-$0$ agents \citep{wright12,wright16}.  
The Spike-Poisson specification of the level distribution allows for precise tuning of the fraction of level-$0$ agents, while also allowing the modeling of higher-level agents.  At the same time, it requires only two parameters, unlike a categorical distribution which requires a number of parameters equal to the maximum level minus one.

\subsection{Data}
\label{sec:data}

We analyzed data from the ten experimental studies summarized in Table~\ref{tbl:datasets}.
In \citet{stahl94} experimental subjects played 10 normal-form games for points, where every point represented a 1\% chance (per game) of winning $\$2.50$.  Participants stood to earn between $\$0.25$ and $\$25.00$ based on their play in the games.
In \citet{stahl95}, subjects played 12 normal-form games, where each point gave a 1\% chance (per game) of winning $\$2.00$. Participants stood to earn between $\$0.00$ and $\$24.00$ based on their play in the games.
In \citet{costagomes98} subjects played 18 normal-form games, with each point of payoff worth 40 cents. However, subjects were paid based on the outcome of only one randomly-selected game. Participants stood to earn between $\$7.84$ and $\$36.16$ based on their play in the games.
\citet{goeree01} presented 10 games in which subjects' behavior was close to that predicted by Nash equilibrium, and 10 other small variations on the same games in which subjects' behavior was \emph{not} well-predicted by Nash equilibrium.  The payoffs for each game were denominated in pennies.  We included the 10 games that were in normal form.
Participants stood to earn between $\$-1.02$ and $\$23.30$ based on their play in these 10 games.
In \citet{cooper03}, agents played the normal forms of 8 games, followed by extensive form games with the same induced normal forms; we include only the data from the normal-form games.  Payoffs were denominated in 10 cent units.
Participants stood to earn between $\$0.80$ and $\$4.80$ based on their play in the games.
In \citet{haruvy01}, subjects played 15 symmetric $3 \times 3$ normal form games.  The payoffs were points representing a percentage chance of winning \$2.00 for each game.
Participants stood to earn between $\$0.00$ and $\$30.00$ based on their play in the games.
In \citet{costagomes08}, subjects played 14 games, and were paid \$0.15 per point in one randomly-chosen game.
Participants stood to earn between $\$1.83$ and $\$14.13$ based on their play in the games.
In \citet{haruvy07}, subjects played 20 games, again for payoff points representing a percentage chance of winning \$2.00 per game.
 Participants stood to earn between $\$1.05$ and $\$17.40$ based on their play in the games.
\citet{stahl08} presented new data on 15 games that contain strategies that are dominated in ways that are ``obvious'' to varying degrees, again for percentage chances of winning \$2.00 per game.
Participants stood to earn between $\$0.00$ and $\$17.55$ based on their play in the games.
Finally, in \citet{rogers09}, subjects played 17 normal-form games, with payoffs denominated in pennies.
Participants stood to earn between $\$2.31$ and $\$13.33$ based on their play in the games.

We represent the data for each game $G_i$ as a pair $(G_i, \{a_{ij}\})$ containing the game itself and a set of observed actions in the game.
All games had two players, so each single play of a game generated two observations.  We built one such dataset for each study, as listed in Table~\ref{tbl:datasets}.
%
We also constructed a combined dataset, dubbed \noun{All10}, containing data from all the datasets.
The datasets contained very different numbers of observations, ranging from 400 \citep{stahl94} to 2992 \citep{cooper03}.
To ensure that each fold had approximately the same population of subjects and games, we evaluated \noun{All10} using \emph{stratified} cross-validation, ensuring that the number of games from each source dataset was approximately equal in each partition element.
\klb{This is a little bit informal. Why is it only approximately equal? ``we performed the game partitioning and selection process separately'' doesn't specify the process quite explicitly.}
\jrw{They are approximately equal because most datasets don't have a number of games that is an exact multiple of 10.}
%
We also renormalized all games so that their payoffs were expressed in expected cents. This is important because the precision parameter for quantal response is not \emph{scale invariant}: the correct value of $\lambda$ can differ depending upon the units in which payoffs are expressed.  
As described earlier, in some datasets, payoff points were worth a certain number of cents; in others, points represented percentage chances of winning a certain sum, or were otherwise in expected units.
Table~\ref{tbl:datasets} lists the number of expected cents that we deemed each payoff point to be worth for the purpose of this normalization.

\begin{table}[tb]\centering
  \caption[Names and contents of each dataset.]{
  Names and contents of each dataset.  Units are in expected value, in US dollars.}
  \label{tbl:datasets}
  {
  \begin{tabular}{lp{2in}lll}
      \toprule
        Name         & Source                & Games & $n$ & Units\\
      \midrule       
        \noun{SW94}  & \citet{stahl94}       &  10   &  400  &  \$0.025 \\
        \noun{SW95}  & \citet{stahl95}       &  12   &  576  &  \$0.02  \\
        \noun{CGCB98}& \citet{costagomes98}  &  18   & 1566  &  \$0.022 \\
        \noun{GH01}  & \citet{goeree01}      &  10   &  500  &  \$0.01  \\
        \noun{CVH03} & \citet{cooper03}      &   8   & 2992  &  \$0.10  \\
        \noun{HSW01} & \citet{haruvy01}      &  15   &  869  &  \$0.02  \\
        \noun{HS07}  & \citet{haruvy07}      &  20   & 2940  &  \$0.02  \\
        \noun{CGW08} & \citet{costagomes08}  &  14   & 1792  &  \$0.0107 \\
        \noun{SH08}  & \citet{stahl08}       &  18   & 1288  &  \$0.02  \\
        \noun{RPC08} & \citet{rogers09}      &  17   & 1210  &  \$0.01  \\
      \midrule
        \noun{All10} &  Union of above       & 142   & 13863 &  per source \\
      \bottomrule\\
    \end{tabular}
  }
\end{table}

\subsection{Methods}
\label{sec:methods}

Models cannot be evaluated unless their parameters are instantiated; we set parameters to values that maximized the likelihood the model assigned to a training dataset. We perform this likelihood maximization using the Covariance Matrix Adaptation Evolution Strategy (CMA-ES) algorithm \citep{hansen01}. We then score a given model's performance by the likelihood it assigned to a test dataset, consisting entirely of \emph{games} (and, hence, observations) that were not used for estimating parameters. 
%
Randomly dividing our experimental data into training and test sets introduces variance into the prediction score, since the exact value of the score depends partly upon the random division.  To reduce this variance, we perform 10 rounds of 10-fold \emph{cross-validation}.  Specifically, for each round, we randomly order the games and then divide them into 10 equal-sized parts.  For each of the 10 ways of selecting 9 parts from the 10, we compute the maximum likelihood estimate of the model's parameters based on the observations associated with the games of those 9 parts.  We then determine the likelihood of the remaining part given the prediction. We call the average of this quantity across all 10 parts the \emph{cross-validated likelihood}.  The average (across rounds) of the cross-validated likelihoods is distributed according to a Student's-$t$ distribution \citeeg{witten00}.  We compare the predictive power of different behavioral models on a given dataset by comparing the average cross-validated likelihood of the dataset under each model.  We say that one model predicts significantly better than another when the $95\%$ confidence intervals for the average cross-validated likelihoods do not overlap.

In Section~\ref{sec:l0-parameters}, we analyze posterior distributions over the parameters of the models we consider.
To compute these posteriors, we used \emph{Metropolis-Hastings sampling} to estimate the posterior distribution.  The Metropolis-Hastings algorithm is a Markov Chain Monte Carlo (MCMC) algorithm \citeeg{robert04} that computes a series of values from the support of a distribution.
Although each value depends upon the previous value, the values are distributed as if from an independent sample of the distribution after a sufficiently large number of iterations.
MCMC algorithms (and related techniques, e.g., annealed importance sampling \citep{neal01}) are useful for estimating multidimensional distributions for which a closed form of the density is unknown.  They require only that a value \emph{proportional} to the true density be computable (i.e., an unnormalized density).  This is precisely the case with the models that we seek to estimate.
%
We used a flat prior for all parameters.
}

\section{Level-0 Model}
\label{sec:l0-models}

In this section we present the components from which we will construct models for computing level-$0$ distributions of play.  
%
We first describe features computed for each of an agent's actions, followed by options for combining feature values to obtain a level-$0$ prediction.

\subsection{Level-0 Features}
\label{sec:features}

Most applications of iterative models specify that level-$0$ agents choose their actions uniformly, thus implicitly identifying nonstrategic behavior with uniform randomization.
The core idea of this \paper{} is that nonstrategic behavior need not be uniform. How then might a nonstrategic agent behave?
We argue that agents consider simple rules (\emph{features}) that recommend one or more actions, to greater or lesser degrees.  We consider both binary features with range $\{0,1\}$ and real-valued features with range $\R^+$.

To be eligible for inclusion in our level-$0$ specification, we require that features not depend on beliefs about how other agents will attempt to maximize their own utility functions.
E.g., the maxmax payoff feature (see below) could be interpreted as a belief that the other agents will act in such a way that the level-$0$ agent can maximize its own payoff; however, this belief does not take the other agents' utility function into account at all.
The minmin unfairness feature (see below) could be interpreted as a belief that the other agents will act to minimize unfairness; however, it does not model the other agents as attempting to maximize their own utility at all.

We restrict our attention to features that can be computed directly from the normal form of the game, and which do not depend on presentation details such as the units in which payoffs are expressed or the order in which actions are presented.  This allows for more accurate analysis of strategic models, even when details of presentation are unknown or not yet known. We do not claim that the features that we investigated comprise an exhaustive list of factors that could influence nonstrategic agents' actions.
\thesisonly{In Chapter~\ref{ch:gamenet}, we will consider a more automated but less interpretable procedure for learning nonstrategic features.}

For each feature, we briefly describe its motivation and then formally define it. Many of our features have been investigated in both the classical and behavioral game theory literature in other contexts.  In particular, the maxmax payoff, maxmin payoff, and maxmax welfare features correspond to the \emph{Optimistic}, \emph{Pessimistic}, and \emph{Altruistic} nonstrategic types in \citet{costagomes01}.  Other features, such as the max-symmetric feature, were influenced by introspection about paradigmatic games such as the Traveler's Dilemma.

For each feature, we define both a binary version and a real-valued version.
Unlike a binary feature, where a criterion must be maximized in order to be recommended, with a real-valued feature an action will be recommended to the \emph{degree} that it maximizes a criterion.  This addresses the intuition that two very high payoff actions may both be attractive, even if one offers marginally higher payoff than the other.

Some real-valued features represent quantities that an agent would wish to minimize, rather than maximizing.
We apply the $\text{inv}$ transformation to these features, where $\text{inv}$ is defined differently depending upon how features will be combined.
If feature values will be combined linearly, then $\text{inv}(x) = 1/x$.
If feature values will be combined with a logit function, then $\text{inv}(x) = -x$.

\paragraph{Maxmin payoff.}
A maxmin action for agent $i$ is the action with the best worst-case guarantee.  That is,
\[f^{\text{maxmin}}(a_i) = \begin{cases}
1 &\text{if } a_i \in \argmax_{a'_i \in A_i} \min_{a_{-i} \in A_{-i}} u_i(a'_i, a_{-i}), \\
0 &\text{otherwise.}
\end{cases}
\]
This is the safest single action to play against a hostile agent.\footnote{Often, a mixed strategy will be safer still against a hostile agent.  However, in this application we are not actually trying to find a safest strategy for the agent.  Rather, we are trying to specify features of individual actions that might make them attractive to nonstrategic agents.}  The real-valued version of this feature returns the worst-case payoff for an action:
\[f^{\text{min}}(a_i) = \min_{a_{-i} \in A_{-i}} u_i(a_i, a_{-i}).\]

\paragraph{Maxmax payoff.}
In contrast, a maxmax action for agent $i$ is the action with the best best case.  That is,
\[f^{\text{maxmax}}(a_i) = \begin{cases}
1 &\text{if } a_i \in \argmax_{a'_i \in A_i} \max_{a_{-i} \in A_{-i}} u_i(a'_i, a_{-i}), \\
0 &\text{otherwise.}
\end{cases}
\]
An agent who aims to maximize his possible payoff will play a maxmax action.
The real-valued version of this feature returns the best-case payoff for an action:
\[f^{\text{max}}(a_i) = \max_{a_{-i} \in A_{-i}} u_i(a_i, a_{-i}).\]

\paragraph{Minimax regret.}
\citet{savage51} proposed the \emph{minimax regret} criterion for making decisions in the absence of probabilistic beliefs.  In a game theoretic context, it works as follows.  For each action profile, an agent has a possible \emph{regret}: how much more utility could the agent have gained by playing the best response to the other agents' actions?  Each of the agent's actions is therefore associated with a vector of possible regrets, one for each possible profile of the other agents' actions.  A minimax regret action is an action whose maximum regret (in the vector of possible regrets) is minimal.  That is, if
\[r(a_i, a_{-i}) = u_i(a_i, a_{-i}) - \max_{a^*_i \in A_i} u_i(a^*_i, a_{-i})\]
is the regret of agent $i$ in action profile $(a_i, a_{-i})$, then
\[f^{\text{mmr}}(a_i) = \begin{cases}
1 &\text{if } a_i \in \argmin_{a'_i \in A_i} \max_{a_{-i} \in A_{-i}} r(a_i, a_{-i}),\\
0 &\text{otherwise.}
\end{cases}
\]
The real-valued version of this feature returns the worst-case regret for playing an action:
\[f^{\text{mr}}(a_i) = \text{inv}\left[\max_{a_{-i} \in A_{-i}} r(a_i, a_{-i})\right].\]
Higher max regret is less desirable than lower max regret, explaining our use of the $\text{inv}$ transformation.

\paragraph{Minmin unfairness.}
Concern for the fairness of outcomes is a common feature of human play in strategic situations, as has been confirmed in multiple behavioral studies, most famously in the Ultimatum game \citep{thaler88,camerer95}.
Let the unfairness of an action profile be the difference between the maximum and minimum payoffs among the agents under that action profile: \[d(a) = \max_{i,j \in N} u_i(a) - u_j(a).\]
Then a \emph{fair} outcome  minimizes this difference in utilities.  The \emph{minmin unfairness} feature selects every action which is part of a minimally unfair action profile.
\[f^{\text{fair}}(a_i) = \begin{cases}
1 &\text{if } a_i \in \argmin_{a'_i \in A_i} \min_{a_{-i} \in A_{-i}} d(a'_i, a_{-i}), \\
0 &\text{otherwise.}
\end{cases}
\]
The real-valued version of this feature returns the minimum unfairness that could result from playing a given action:
\[f^{\text{unfair}}(a_i) = \text{inv}\left[\min_{a_{-i} \in A_{-i}} d(a_i, a_{-i})\right].\]
Unfairness is a quantity to be minimized, so we apply the $\text{inv}$ transformation.

\paragraph{Max symmetric.}
\klb{I simplified this definition by removing the $u^{\text{symm}}$ notation. Revert it if you like.}%
\jrw{Makes sense to me; I added \ldots to make it apply to $n$-player games.}%
People often reason about what would happen if the other agent acted as they did.\footnote{This is a concept that only applies to symmetric games, in which agents have identical action sets, and each agent's payoff matrix is the transpose of the other.} 
A max-symmetric action is simply the best such action:
\[f^{\text{maxsymm}}(a_i) = \begin{cases}
1 &\text{if } a_i \in \argmax_{a'_i \in A_i}  u(a'_i, \ldots, a'_i), \\0 &\text{otherwise.}
\end{cases}
\]
The real-valued version of this feature returns the symmetric payoff of an action:
\[f^{\text{symm}}(a_i) = u(a_i, \ldots, a_i).\]

\paragraph{Maxmax welfare.}
\klb{I wonder if maxmax welfare is a better name for this. I think of efficiency as a name for the optimum, not the quantity.}%
\jrw{Sure, makes sense to me.}
Finally, one reason that a nonstrategic agent might find an action profile desirable is that it produces the best overall benefit to the pair of agents.  The \emph{maxmax welfare} feature selects every action that is part of some action profile that maximizes the sum of utilities:
\[f^{\text{efficient}}(a_i) = \begin{cases}
1 &\text{if } a_i \in \argmax_{a'_i \in A_i} \max_{a_{-i} \in A_{-i}} \sum_{j \in N} u_j(a'_i, a_{-i}), \\
0 &\text{otherwise.}
\end{cases}
\]
The real-valued version of this feature returns the maximum welfare that could result from playing a given action:
\[f^{\text{welfare}}(a_i) = \max_{a_{-i} \in A_{-i}} \sum_{j \in N} u_j(a_i, a_{-i}).\]


\subsection{Combining Feature Values}
\label{sec:functional-forms}

Once a set of features have been computed for each of a set of actions, their values must be combined to yield a single distribution over actions.  There is an infinite number of ways to perform such a combination.  We considered two functional forms, inspired by linear regression and logit regression respectively.

Both specifications accept a set of features and a set of weights.
Let $F$ be a set of features mapping from an action to $\R^+$.
For each feature $f \in F$, let $w_f \in [0,1]$ be a weight parameter.
Let $\sum_{f \in F} w_f \le 1$, and let $w_0 = 1 - \sum_{f \in F} w_f$.

The first functional form produces a level-$0$ prediction over actions for a given agent by taking a weighted sum of feature outputs for each action and then normalizing to produce a distribution.
\begin{definition}[Weighted linear level-$0$ specification]
    The \emph{weighted linear level-$0$ specification} predicts the following distribution of actions for level-$0$ agents:
	\[\pi^{\text{linear},F}_{i,0}(a_i) = \frac{w_0 + \sum_{f \in F} w_ff(a_i)}
	{\sum_{a'_i \in A_i} \left[w_0 + \sum_{f \in F} w_ff(a'_i)\right]}.
	\]
\end{definition}

The second functional form assigns a level-$0$ probability proportional to the exponential of a weighted sum of feature values.  
\begin{definition}[Logit level-$0$ specification]
    The \emph{logit level-$0$ specification} predicts the following distribution of actions for level-$0$ agents:
	\[\pi^{\text{logit},F}_{i,0}(a_i) = \frac{\exp(w_0 + \sum_{f \in F} w_ff(a_i))}
	{\sum_{a'_i \in A_i} \exp(w_0 + \sum_{f \in F} w_ff(a_i))}.
	\]
\end{definition}

\subsection{Feature Transformations}
\label{sec:feature-transformations}

In addition to two functional forms for combining the feature values, we also evaluated two transformations to feature values.  These transformations may be applied to each feature value before they are weighted and combined.

The first transformation zeroes out features that have the same value for every action, which we call \emph{uninformative}.  The intuition behind this transformation is that  informative features should have a greater influence on the prediction precisely when the other features are less informative.

\begin{definition}[Informativeness feature transformation]
    A feature $f$ is \emph{informative} in a game $G$ if there exists $a'_i, a''_i \in A_i$ such that $f(a'_i) \ne f(a''_i)$.
    The informativeness transformation $I(f)$ of a feature $f$ returns the feature's value when it is informative, and zero otherwise:
    \[I(f)(a_i) = \begin{cases}
	f(a_i) &\text{if } \exists a'_i,a''_i \in A_i:\; f(a'_i) \ne f(a''_i), \\
	0 &\text{otherwise.}
	\end{cases}
    \]
\end{definition}

The second transformation normalizes feature values to $[0,1]$.
This limits the degree to which one real-valued feature can overwhelm other features.

\begin{definition}[Normalized activation feature transformation]
    The normalized activation transformation $N(f)$ constrains a feature $f$ to take nonnegative values that sum to 1 across all of a game's actions:
    \[N(f)(a_i) = \frac{f(a_i)}{\sum_{a'_i \in A_i} f(a'_i)}.\]
\end{definition}

\section{Model Selection}
\label{sec:model-selection}

We took two approaches to constructing a model from the candidate features, functional forms, and transformations described in the previous section.  First, we performed \emph{forward selection} of binary features, using a linear functional form and informativeness transformation.  We chose this functional form based on a manual evaluation we performed in the conference version of this \paper{} \citep{wright14level}, in which a linear functional form and normalized-activation- and informativeness-transformed binary features yielded good performance.
Second, we performed \emph{Bayesian optimization} to automatically evaluate combinations from the full set of candidate features, functional forms, and transformations.

\subsection{Forward Selection}
\label{sec:forward-selection}

\begin{figure}[t]\centering
    \includegraphics{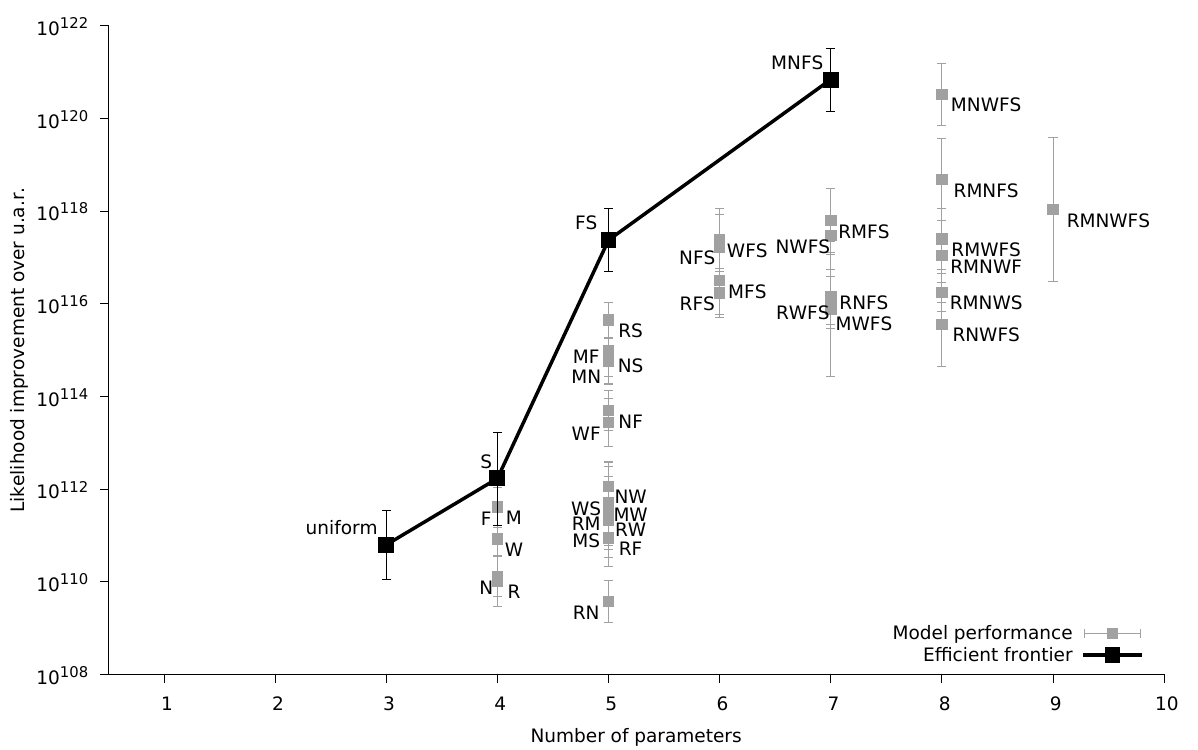}
    \caption[Prediction performance with binary features]{
        Prediction performance with 95\% confidence intervals for Spike-Poisson QCH extended by binary features.
        Points are labeled by a code indicating which features were included:
        (M) maxmax payoff; (N) maxmin payoff; (R) minmax regret; (W) maxmax welfare; (F) minmin unfairness; (S) max symmetric.}
    \label{fig:forward-selection}
\end{figure}
\begin{figure}[t]\centering
    \includegraphics{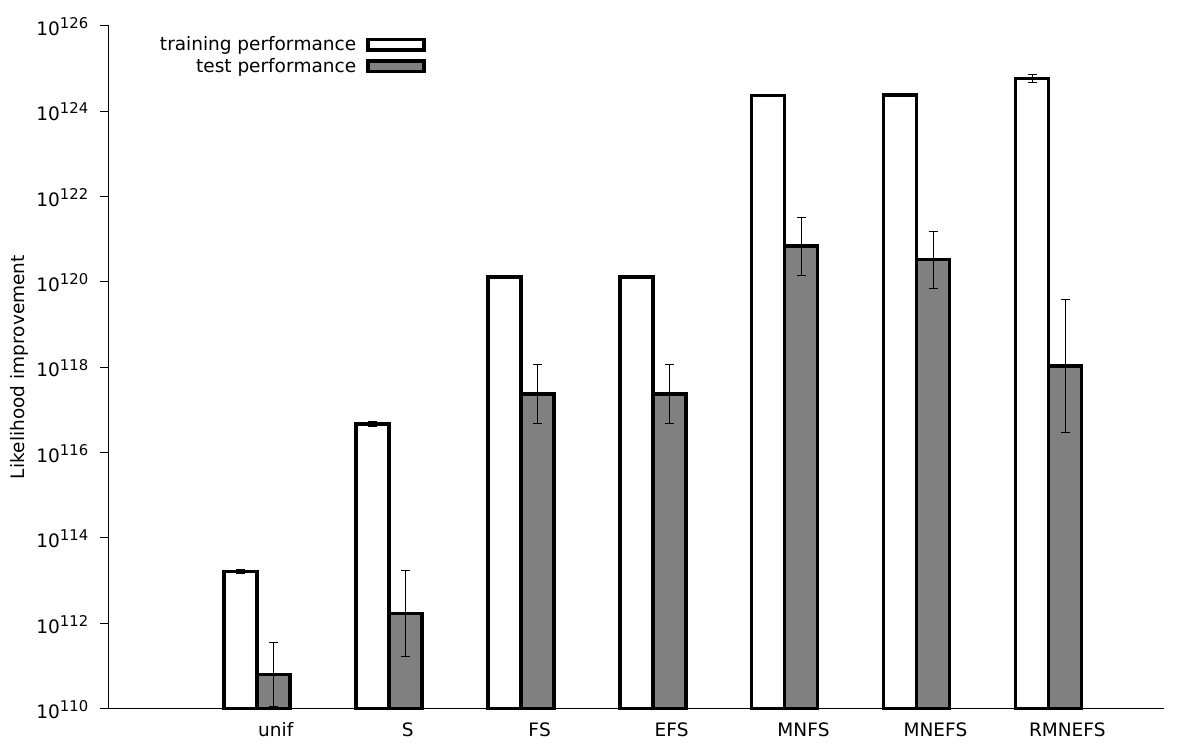}
    \caption[Training and test performance with binary features]{
        Training and test performance with 95\% confidence intervals for Spike-Poisson QCH extended by binary features, for the best performing set of features at each number of features.}
    \label{fig:l0-overfitting}
\end{figure}

We performed forward selection using the following procedure.
We evaluated the test performance of the Spike-Poisson QCH model, extended by a linear, normalized-activation- and informativeness-transformed level-$0$ model using every one- and two-element subset of the binary features from Section~\ref{sec:features}.  The best such model used the minmin unfairness and max symmetric features.  We then evaluated every combination of features that contained those two features. The results are shown in Figure~\ref{fig:forward-selection}.

The best performing linear model found by forward selection contained four features: maxmax payoff, maxmin payoff, minmin unfairness, and max symmetric.  We will refer to this model henceforth as \linear4.  Adding further features did not improve prediction performance.  Figure~\ref{fig:l0-overfitting} shows the training and test performance for the best-performing model at each number of features.  Notice that the training performance increased with every additional feature, whereas after the four-feature model the test performance decreased.  This confirms that overfitting was the cause of the performance decrease.\footnote{It might seem obvious that a drop in test performance for more general models must imply overfitting.  However, it could also indicate \emph{underfitting}, where we simply do a worse job of optimizing the more complex models.  It is this latter possibility that Figure~\ref{fig:l0-overfitting} rules out.}

\subsection{Bayesian Optimization}
\label{sec:bayesian-optimization}

We performed Bayesian optimization using SMAC \citep{hutter10,hutter11,hutter12}, a software package for optimizing the configuration of algorithms.  SMAC evaluates each configuration on a randomly-chosen \emph{instance} (i.e., input to the algorithm); it then updates a random forest model of predicted performance for configurations.  It determines which configurations to evaluate based on the performance model.

We ran 16 parallel SMAC processes for 1200 hours each.  The processes shared the results of each run they performed.
In the context of choosing a level-$0$ specification, a configuration is a set of features, a set of feature transformations, and a functional form choice.  An instance is a \emph{subfold}: a seed used to randomly divide the \noun{All10} dataset into 10 folds; an index indicating which fold is the test fold; and an index indicating which subdivision of the training folds to use as the validation fold.
The specified configuration was trained on the training data minus the validation fold; the performance of the trained model on the validation fold was then output to SMAC.  The test fold was ignored.  In this way, we attempted to avoid overfitting our dataset during model selection, by never using the test fold that we used for our final model evaluations to evaluate candidate configurations. 

\begin{figure}[t]\centering
    \includegraphics{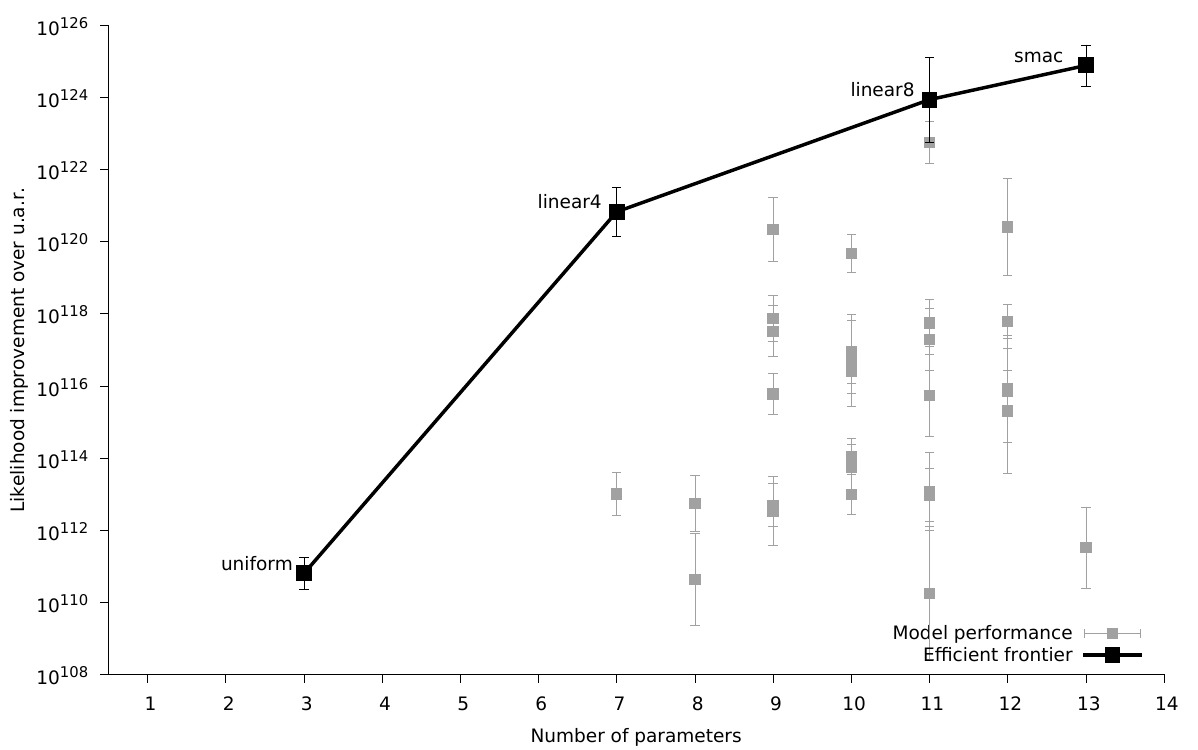}
    \caption[Prediction performance for Bayesian optimization incumbents]{
        Prediction performance with 95\% confidence intervals for Spike-Poisson QCH extended by features, functional form, and feature transformations selected by Bayesian optimization.  We show only models that were at any point ``incumbent'' (i.e., the best found by SMAC at some point in time).}
    \label{fig:incumbents}
\end{figure}

Figure~\ref{fig:incumbents} shows the results of the Bayesian optimization process.
The best-performing model found by SMAC was a $13$-parameter model that contained the same four binary features as \linear4, plus \emph{all} of the real-valued features described in Section~\ref{sec:features}.  It combined the features linearly, with both the normalized-activation and informativeness transformations.  We refer to this model as \smac{}.

We were intrigued that SMAC's best-performing model was essentially \linear4, augmented by real-valued features.
We hypothesized that \linear4 augmented by real-valued versions of its four binary features only (i.e., excluding the welfare and max-regret features) would perform as well or better as \smac{}.
This model, which we refer to as \linear8, was not checked by SMAC, and so we checked it manually.
As shown in Figure~\ref{fig:incumbents}, there was no significant difference between the performance of \linear8 and \smac{} (although \smac{} did insignificantly outperform \linear8), even though \linear8 has two fewer features.
For the remainder of the \paper{}, we  focus our attention on the \linear4 and \linear8 models.

\subsection{Extended Model Performance}
\label{sec:performance}

\begin{figure}\centering
    \includegraphics{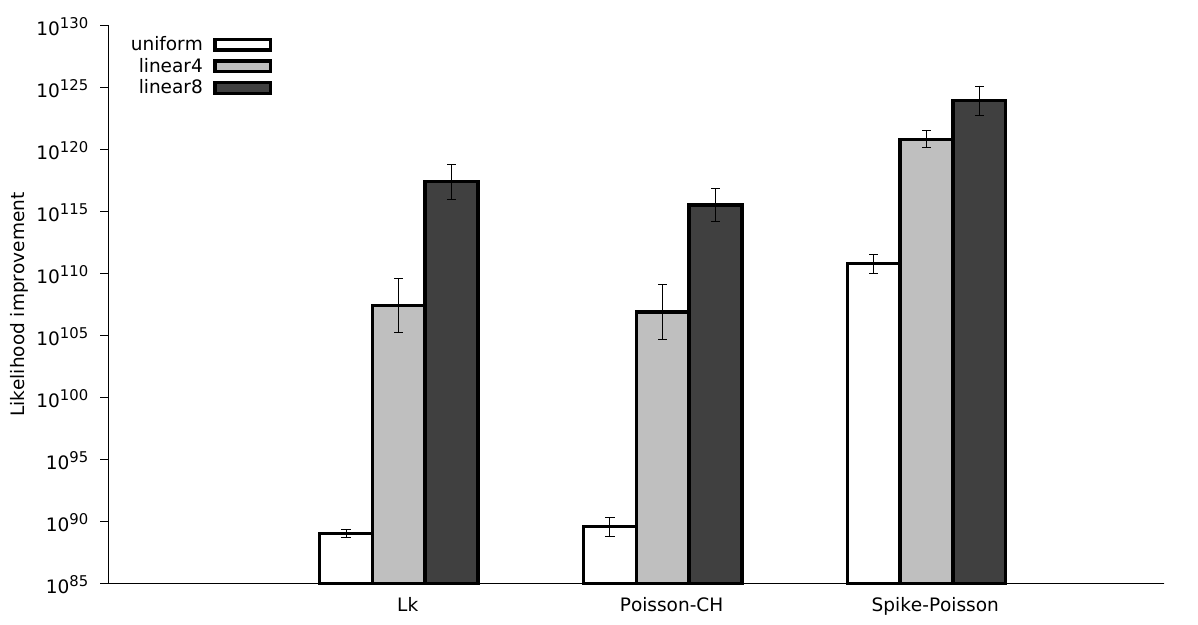}
    \caption[Prediction performance for Poisson-CH, Lk, and Spike-Poisson QCH with \linear8, \linear4, and uniform level-$0$ specifications.]{
        Average likelihood ratios of model predictions to random predictions, with $95\%$ confidence intervals.  Results are shown for three different iterative models (Poisson cognitive hierarchy \citep{camerer04cognitive}, level-$k$ \citep{costagomes01}, and Spike-Poisson quantal cognitive hierarchy
        \thesisonly{[see Section~\ref{sec:spike-poisson}][see Section~\ref{sec:spike-poisson}])}
        \journalonly{\citep{wright16})} using three different level-$0$ specifications (uniform randomization, \linear4 from Section~\ref{sec:forward-selection}, and \linear8 from Section~\ref{sec:bayesian-optimization}).}
    \label{fig:extended}
\end{figure}

We compared the predictive performance of three iterative models using three different specifications of level-$0$ behavior.  The results are displayed in Figure~\ref{fig:extended}.  The $y$-axis gives the average ratio of the cross-validated likelihood of the extended models' predictions divided by the likelihood of a uniform random prediction.
Overall, the \linear4 specification yielded a large performance improvement, both on Spike-Poisson QCH and also on the two other iterative models.  The \linear8 specification yielded an additional, smaller performance improvement.
In fact, the two other iterative models benefited disproportionately from the improved level-$0$ specifications. Spike-Poisson QCH performed better than the other two models under all level-$0$ specifications, but the three models had much more similar (and improved) performance under the \linear4 and \linear8 specifications.
\thesisonly{
Adding the \linear4 level-$0$ model to Lk improved its performance by a factor of about $10^{17}$.
For context, this was nearly as large as the performance gap of $10^{19}$ between Lk and QLk in Chapter~\ref{ch:perf}.
}
This is especially interesting given that the level-$0$ model was selected based solely on the degree to which it improved Spike-Poisson QCH's performance.

\subsection{Parameter Analysis}
\label{sec:l0-parameters}

We now examine and interpret the posterior distributions for some of the parameters of the Spike-Poisson model combined with the \linear4 and \linear8 level-$0$ specifications, following \journalonly{methodology we introduced in prior work \cite{wright12,wright16}.}
\thesisonly{the methodology of Section~\ref{sec:posterior-estimation}.}

\begin{figure}\centering
    \includegraphics[width=0.95\textwidth]{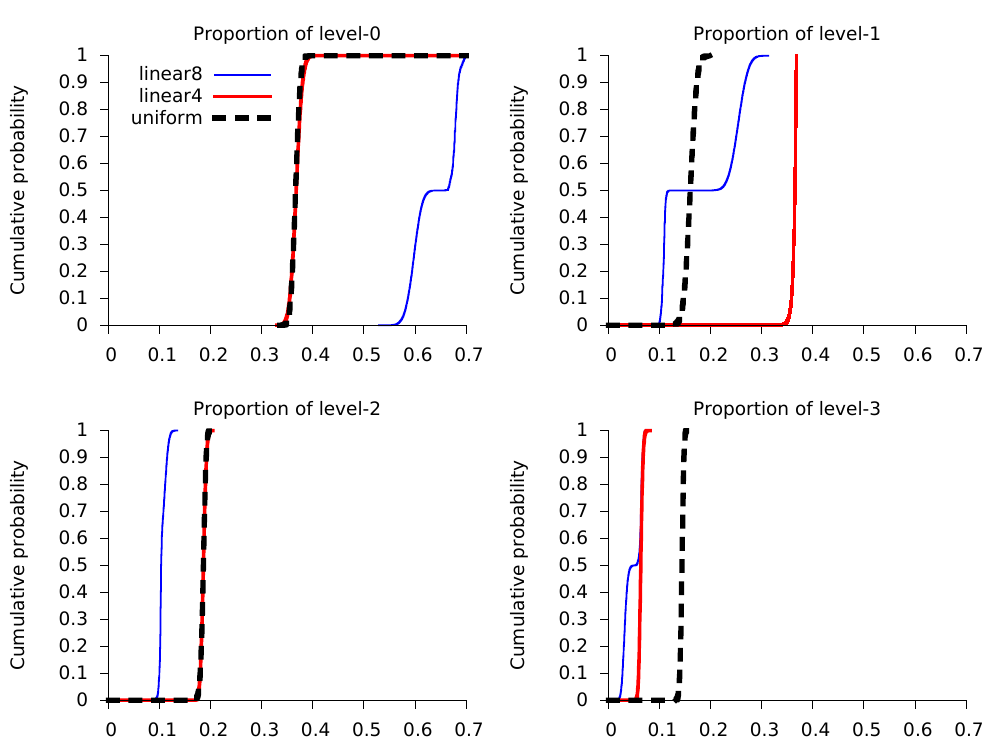}
    \caption[Posterior distributions of levels for \linear4, \linear8, and uniform]{
        Marginal cumulative posterior distributions of levels of reasoning in the \noun{All10} dataset, for Spike-Poisson QCH with \linear8, \linear4, and uniform specifications.}
    \label{fig:cdf-l0-levels}
\end{figure}

Figure~\ref{fig:cdf-l0-levels} shows the marginal posterior distribution for each level in the population (up to level $3$), for each of the \linear4, \linear8, and uniform specifications.
We noticed two effects.  First, the posterior distribution of level-$0$ and level-$2$ agents was essentially identical under the uniform and \linear4 specifications, with medians of $0.37$.  We found this surprising, since the level-$0$ agents behave very differently under the two specifications.  In contrast, the posterior distribution of level-$0$ agents under the \linear8 specification had a median of $0.65$, nearly twice as large.
%
Second, there was a large shift of mass (approximately $0.2$) from agents higher than level-$3$ to level-$1$ agents under the \linear4 specification, and from all nonzero-level agents to level-$0$ under the \linear8 specification.
\thesisonly{
This may indicate that models with a uniform level-$0$ specification were using higher-level agents to simulate a more accurate level-$0$ specification, in much the same way that QLk seemed to be using a low precision-beliefs parameter ($\lambda_{1(2)}$) to simulate a cognitive hierarchy model in Chapter~\ref{ch:parameter-analysis}.
}
\journalonly{
This may indicate that models with a uniform level-$0$ specification were using higher-level agents to simulate a more accurate level-$0$ specification.
}

\begin{figure}\centering
    \includegraphics[width=0.95\textwidth]{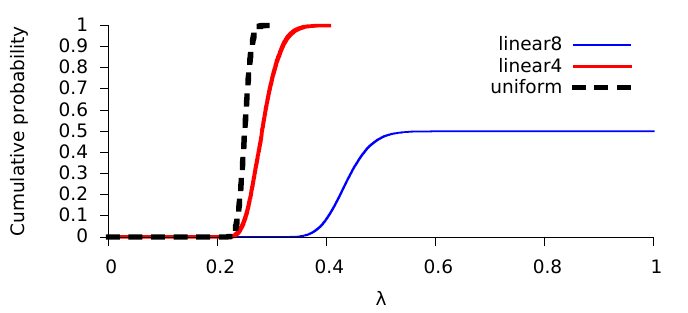}
  \caption[Posterior precision distributions for \linear8, \linear4, and uniform]{
        Marginal cumulative posterior distributions of the precision parameter ($\lambda$) in the \noun{All10} dataset, for Spike-Poisson QCH with \linear8, \linear4, and uniform specifications.}
  \label{fig:cdf-l0-lam}
\end{figure}

The precision parameter was very similar between the uniform and \linear4 specifications.  It was less sharply identified under the \linear4 specification, with mass shifting toward slightly higher precisions. Quantal response serves two purposes: it accounts for the errors of reasoning that people actually make and it also provides the error structure for the model itself.  The higher precision may simply reflect the \linear4 specification's improved accuracy.
Under the \linear8 specification, the precision parameter is not well identified, but has a much higher value with high probability.  This lack of identification may arise from the small role played by nonzero agents in this specification. \klb{That's a good argument, but I'm a bit worried that there's a numerical error or some other problem with \linear8. We've seen in the past that multimodality is a sign of things going wrong, and nearly every blue line has multiple modes. In Fig 6, we don't even get to see past the first mode...}
\jrw{I'm kind of worried about that too; I wonder what the performance and parameters would be if we clamped lambda to infinite (i.e., required best-response)?  QBR adds so much performance normally that the essentially-infinite mode of lambda is pretty worrying.}

\begin{figure}\centering
    \includegraphics{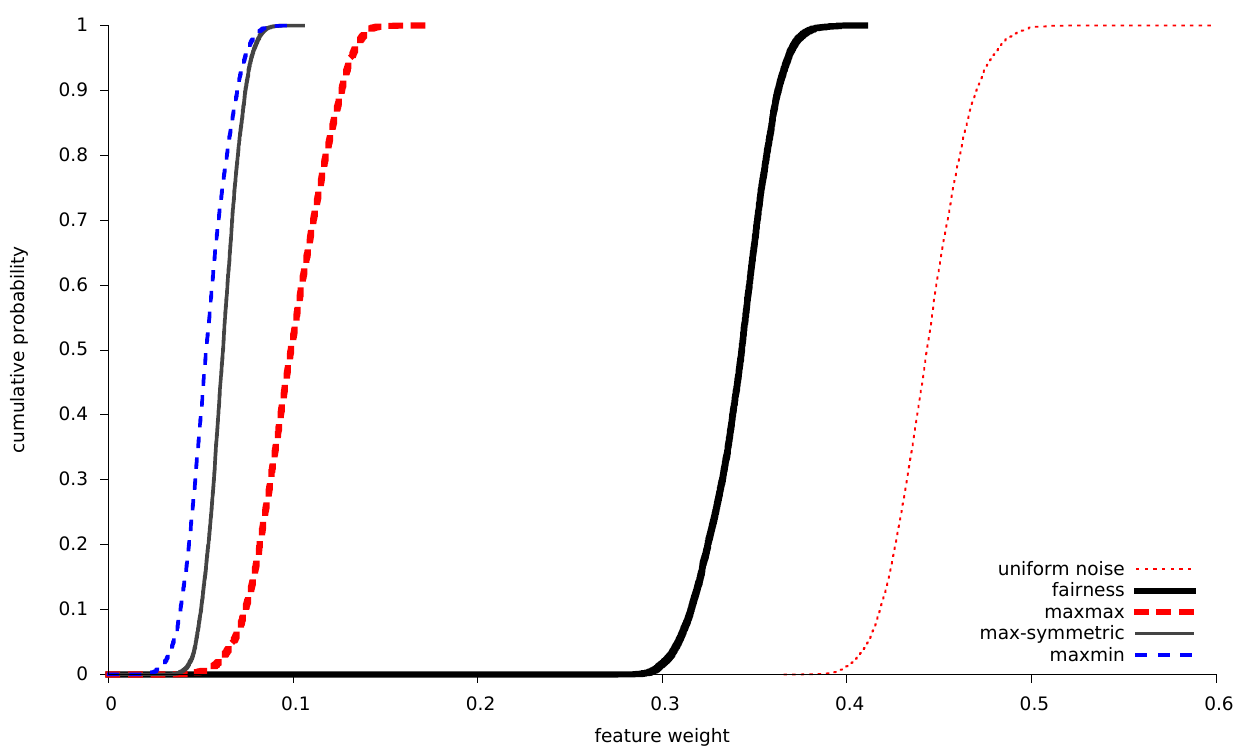}
    \caption[Posterior distribution of weights for \linear4]{
        Marginal cumulative posterior distributions over weight parameters of the \linear4 specification, for Spike-Poisson QCH on the \noun{All10} dataset.  The uniform noise weight is defined implicitly by the other four weights.}
    \label{fig:cdf-linear4-weights}
\end{figure}

\begin{figure}\centering
    \includegraphics{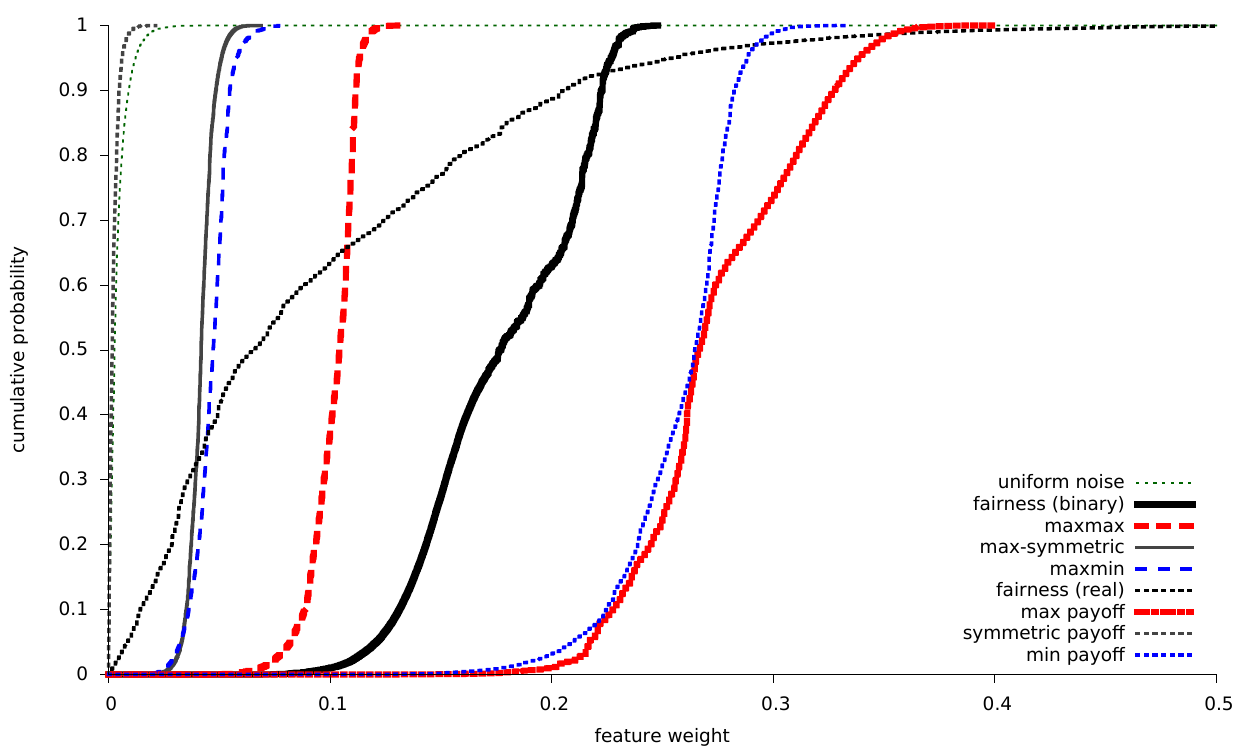}
    \caption[Posterior distribution of weights for \linear8]{
        Marginal cumulative posterior distributions over weight parameters of the \linear8 specification, for Spike-Poisson QCH on the \noun{All10} dataset.  The uniform noise weight is defined implicitly by the other eight weights.}
    \label{fig:cdf-linear8-weights}
\end{figure}

Figures~\ref{fig:cdf-linear4-weights} and~\ref{fig:cdf-linear8-weights} show the marginal posterior distribution for the weights of the \linear4 and \linear8 models respectively on the \noun{All10} dataset.  As with the distribution over levels, the posterior distributions on the weight parameters had modes with very narrow supports, indicating that the data argued consistently for specific ranges of parameter values; the real-valued fairness feature of the \linear8 specification was the exception to this rule.
%
The binary features had broadly similar weights in both the \linear4 and \linear8 specifications: the fairness feature had by far the highest median posterior weight, the maxmax feature had the second-highest weight, and the max-symmetric and maxmin features both had small and essentially identical weights, with very overlapping posterior distributions.
Interestingly, even though the fairness feature was the highest weighted, it was not selected first by the forward selection procedure (max symmetric was selected first).  This likely indicates that fairness is more predictive than other features when it is present, but that it is predictive in fewer games than max symmetric.
\klb{Actually, this suggests an alternate kind of forward selection: selecting the feature that is most useful in the cases where it is informative, not the feature that is most useful overall.}
\jrw{Indeed!  Definitely worth trying out post-groundhog day, when we can compare performance boosts on different datasets.}

The max-payoff and min-payoff real-valued features had very similar posterior weights in the \linear8 specification, with overlapping posterior supports.  These were the highest-weighted features in the \linear8 specification.  The real-valued fairness feature was not well identified.  The symmetric-payoff feature was well-identified and had a very small weight; evidently, the action with the highest symmetric payoff is somewhat salient, but the actual value of the payoff is not salient in itself.

The weight allocated to uniform randomization between the \linear8 and \linear4 specifications is very different; the \linear4 specification allocates nearly half of its weight to uniform randomization, whereas for the \linear8 specification uniform randomization plays almost no part.  This, combined with the strong similarity in the weighting of binary features between the two specifications, suggests that the real-valued features (especially the max and min payoff features) are playing a genuine role in reducing uncertainty.

\section{Related Work}
\label{sec:l0-related}

Almost every study that employs iterative reasoning models of either the level-$k$ or cognitive hierarchy types assumes a uniform distribution of play for level-$0$ agents.  However, there are a few exceptions.
\citet{crawford07} specified truth-telling as the single salient action in first-price auctions.
\citet{crawford07fatal} manually designated certain actions as ``salient'' (based on visual features such as ``leftmost'') in a hide-and-seek game. They then estimated an iterative model with a level-$0$ specification in which level-$0$ agents play salient actions, with the strengths of each action's salience estimated under the assumption that no agent truly plays a level-$0$ distribution.
\citet{arad09} specified a single action of reinforcing all battlefields equally in a Colonel Blotto game as the sole level-$0$ action.
\citet{arad2012money} specified the highest possible request as the level-$0$ action in a money-request game where players receive the amount of money they request, but also receive a relatively large bonus for requesting exactly 1 shekel less than the other player.
\citet{arad2012tennis} manually specified two ``anchor'' strategies for a Colonel Blotto-like game in which players simultaneously assign four representatives to four separate contests in order of the representatives' ability.

In spite of the crucial dependence of iterative models upon the specification of the level-$0$ distribution, few studies have empirically investigated level-$0$ play.  \citet{agranov10} incentivized subjects to choose an action quickly (and then to revise it after thinking) by imposing a randomized time limit when playing the beauty-contest game of \citet{nagel95}.  They hypothesized that early actions represent level-$0$ choices and that later actions represent higher-level choices.  Based on this assumption, they found that level-$0$ behavior did not differ significantly from a uniform distribution.  In contrast, \citet{burchardi12} incentivized players to reveal their reasoning by allowing a one-time simultaneous exchange of messages between teammates playing a beauty-contest game.  Teams of two simultaneously sent each other a single message containing arguments in favor of a given action, and then simultaneously chose an action, with the team's action being chosen randomly from the two choices.  \citeauthor{burchardi12} then classified each argument according to level of reasoning, and extracted the level-$0$ behavior from both level-$0$ and higher-level arguments.  They found that level-$0$ play was significantly different from uniform.  Intriguingly, they also found that the level-$0$ behavior hypothesized by higher-level players was very similar to the level-$0$ behavior that was actually proposed.

\citet{hargreaves14} evaluate the transferability of level-$0$ specifications between three games in which all of the actions are strategically equivalent: a coordination game, a discoordination game, and a hide and seek game.
They argue that any level-$0$ specification based only on the strategic structure of the game must produce an identical level-$0$ behavior for each type of game, since in each game each action is strategically equivalent to every other action.  Based on experimental data, they reject a joint hypothesis that includes an identical distribution of levels for each game and an identical level-$0$ action in each game.\footnote{They initially assume that no level-$0$ agents exist as part of their joint hypothesis. However, their results are robust to the existence of level-$0$ agents.}
This may indicate that framing effects, in addition to strategic considerations, play a role in determining level-$0$ behavior.  It may also indicate that the population distribution of levels varies between games; we are studying this latter possibility in ongoing work.

\section{Conclusions}
\label{sec:conclusions}

This \paper{}'s main contribution is two specifications of level-$0$ behavior that dramatically improve the performance of each of the iterative models we evaluated---level-$k$, cognitive hierarchy, and quantal cognitive hierarchy.  These specifications depend only upon the payoffs of the game, and are thus generally applicable to any domain, even ones in which human intuition gives little guidance about the level-$0$ specification.
%
A linear weighting of four nonstrategic binary features---maxmax payoff, maxmin payoff, minmin unfairness, and max symmetric---improved all three models' performances, with the weaker models (level-$k$ and cognitive hierarchy) improving the most.  Including either or both of the remaining two binary features caused degradations in prediction performance due to overfitting.  Fairness was the feature with by far the greatest weight.  Including real-valued versions of the binary features further improved prediction performance for all three models at the expense of nearly doubling the dimensionality of the level-$0$ specification.

Conventional wisdom in the economics literature says that level-$0$ agents exist only in the minds of higher level agents; that is, that a level-$0$ specification acts solely as a starting point for higher level agents' strategies.  Our results argue against this point of view: the best performing model estimated that more than a third of the agents were level-$0$.  These results are strong evidence that nonstrategic behavior is an important aspect of human behavior, even in strategic settings.  Further refining our understanding of nonstrategic behavior is an important direction for future work, both for the factors that are considered by nonstrategic agents, and for the details of incorporating these factors into predictive behavioral models.

\journalonly{
\section*{Acknowledgements}
This work was funded in part by a Canada Graduate Scholarship from the Natural Sciences and Engineering Research Council of Canada and a Four Year Fellowship from the University of British Columbia.
}


\vskip 0.2in
\bibliography{journal-level0}
\bibliographystyle{theapa}

\end{document}